\documentclass[11pt,a4paper]{emulateapj}
\usepackage{amsmath}
\usepackage{multirow}

\begin{document}

\title{capturing the 3D motion of an infalling galaxy via fluid dynamics}

\author{Yuanyuan Su\altaffilmark{$\ddagger$1}}
\author{Ralph P.\ Kraft\altaffilmark{1}}
\author{Paul E.\ J.\ Nulsen\altaffilmark{1}}
\author{Elke Roediger\altaffilmark{2}}
\author{William R.\ Forman\altaffilmark{1}}
\author{Eugene Churazov\altaffilmark{3}}
\author{Scott W.\ Randall\altaffilmark{1}}
\author{Christine Jones\altaffilmark{1}}
\author{Marie E.\ Machacek\altaffilmark{1}}
\affil{$^1$Harvard-Smithsonian Center for Astrophysics, 60 Garden Street, Cambridge, MA 02138, USA}
\affil{$^2$E.A. Milne Centre for Astrophysics, Department of Physics and Mathematics, University of
Hull, Hull, HU6 7RX, United Kingdom}
\affil{$^3$Max Planck Institute for Astrophysics, Karl-Schwarzschild-Str. 1, 85741, Garching, Germany}

\altaffiltext{$\dagger$}{Email: yuanyuan.su@cfa.harvard.edu}

\begin{abstract}

The Fornax Cluster is the nearest ($\leq20$\,Mpc) galaxy cluster in the southern sky. NGC~1404 is a bright elliptical galaxy falling through the intracluster medium of the Fornax Cluster. The sharp leading edge of NGC~1404 forms a classical ``cold front" that separates 0.6 keV dense interstellar medium and 1.5 keV diffuse intracluster medium. We measure the angular pressure variation along the cold front using a very deep (670\,ksec) {\sl Chandra} X-ray observation. We are taking the classical approach -- using stagnation pressure to determine a substructure's speed -- to the next level by not only deriving a general speed but also directionality which yields the complete velocity field as well as the distance of the substructure directly from the pressure distribution. We find a hydrodynamic model consistent with the pressure jump along NGC~1404's atmosphere measured in multiple directions. The best-fit model gives an inclination of 33$^{\circ}$ and a Mach number of 1.3 for the infall of NGC~1404, in agreement with complementary measurements of the motion of NGC 1404. Our study demonstrates the successful treatment of a highly ionized ICM as ideal fluid flow, in support of the hypothesis that magnetic pressure is not dynamically important over most of the virial region of galaxy clusters.

\end{abstract}

\section{\bf Introduction}

The standard cosmology model, of a universe dominated by dark energy and cold dark matter ($\Lambda$CDM), directly predicts hierarchical structure formation. 
Galaxy clusters, the largest gravitationally collapsed systems in the Universe, are the manifestations. 
The assembly of galaxy clusters is driven by the merger of subclusters, the accretion of galaxy groups, and, most frequently, the infall of dark matter, gas, and galaxies. The growth of galaxy clusters is continuous; truly symmetric and undisturbed clusters are rare.
X-ray emitting hot gas, the intracluster medium (ICM), fills the entire volume of galaxy clusters, and records the cluster formation history.
The {\sl Chandra} X-ray Observatory, with its superb spatial resolution, 
has revealed the ubiquitous presence of ``cold fronts" in the ICM. They are sharp interfaces separating low entropy gas (cold and dense) and high entropy gas (hot and diffuse) (see Markevitch \& Vikhlinin 2007 for a review).  
Cold fronts that are induced by the infall of galaxies or subclusters are often observed at the leading edge of infalling objects (Abell~3667--Vikhilinin et al.\ 2001; M86--Forman et al.\ 1979; Randall et al.\ 2008; NGC~1400--Su et al.\ 2014; M49--Kraft et al.\ 2011). In the case of supersonic motion, bow shocks ahead of infalling objects are also expected.
An accurate determination of the infall geometry is critical for predicting observable shock features in both X-ray and radio, potentially shedding light on how electrons are accelerated and how the ICM is energized.

The flow of the ICM on macroscopic scales can be approximated as an ideal fluid as long as the thermal pressure greatly exceeds the magnetic pressure (e.g., ZuHone \& Roediger 2016).
Our knowledge of fluid dynamics has been widely used to infer cluster gas motion. 
One of the best known examples is the application of the Rankine-Hugoniot jump conditions: 
the gas properties on both sides of a shock wave in a one-dimensional flow can be used to infer the 
infalling speed of a substructure 
(e.g, Su et al.\ 2016; Vikhlinin et al.\ 2001; Markevitch et al.\ 2002). 
Moreover, cosmological applications of galaxy clusters require accurate measurements of cluster masses, which also rely on the hydrostatic approximation of ICM (e.g., Buote et al.\ 2016).

The properties of ICM can be best studied with nearby clusters and their bright member galaxies. The Fornax Cluster, centered on the bright early-type galaxy NGC~1399, is the nearest cluster in the southern sky. It has an ICM temperature of $\approx$1.5 keV and a virial radius of $r_{\rm vir}\footnote{$r_{500}$=0.391 (kT/keV)$^{0.63}$ h${_{70}}^{-1} = 500$\,kpc (Finoguenov et al.\ 2006). $r_{\rm vir}\approx r_{200}\approx 1.5\,r_{500}$ (Yang et al.\ 2009).}\approx 750$\,kpc. 
A {\sl Chandra} mosaic image of the Fornax Cluster is shown in Figure~\ref{fig:fox}.
The cluster center appears to have prominent sloshing edges in the east-west direction. 
NGC~1404 lies at a radius of 12$^{\prime}$ (65\,kpc) to the southeast of NGC~1399. The atmosphere of NGC~1404 is substantially cooler and denser than the ambient ICM; its leading edge forms a classical ``cold front". The Fornax Cluster has been a favorite target for several generations of X-ray telescopes, with a particular focus on NGC~1404 (e.g., 
Jones et al.\ 1997; Buote \& Fabian 1998; Paolillo et al.\ 2002; Buote 2002; Scharf et al.\ 2005; Machacek et al.\ 2005; Murakami et al.\ 2011).
From {\sl ROSAT} observations, Jones et al.\ (1997) suggested NGC~1404 was being stripped due to its infall towards NGC~1399. This scenario was later confirmed by a 134\,ksec {\sl Chandra} observation on NGC~1404 which revealed a sharp leading edge in the direction of NGC~1399 and an extended gaseous tail trailing behind (Machacek et al.\ 2005).

This is the first paper of a series on a very deep (670 ksec) {\sl Chandra} observation of NGC~1404. In this paper, we focus on its macroscopic dynamics. In subsequent papers, we will present its microscopic transport phenomena, properties of the interstellar medium (ISM), and our simulation tailored to the specific merging scenario of NGC~1404. 
%We assume a cosmology with $H_0=75$\,km\,s$^{-1}$\,Mpc$^{-1}$, $\Omega_{\Lambda}=0.7$, and $\Omega_m=0.3$. 
For NGC~1404 and the Fornax Cluster, we adopt a redshift of $z=0.00475$ (the redshift of the central dominant galaxy NGC~1399) from the NASA/IPAC Extragalactic Database (NED) and a luminosity distance of 19 Mpc ($1^{\prime} = 5.49$ kpc) taken from Paolillo et al.\ (2002). 
We describe the observations and data reduction in \S2, and we report results in \S3.
The implications of our results are discussed in \S4, and our main conclusions are summarized in \S5. Uncertainties reported in this paper are at 1$\sigma$ unless stated otherwise.

\section{\bf observations and data reduction}

\subsection{Chandra}
 
The {\sl Chandra} observations used in the analyses presented here are listed in Table~1. A total exposure of 1\,Ms on the Fornax Cluster was included, 670 ksec of which are focused on NGC~1404.
We used {\sl CIAO}~4.8 and {\sl CALDB}~4.6.9 to process and reduce the {\sl Chandra} data. All the observations were reprocessed from level 1 events using the {\sl CIAO} tool {\tt chandra\_repro} such that the latest, consistent calibrations were used. 
%Only events with grades 0, 2, 3, 4, and 6 were included. We also removed bad pixels, bad columns, and node boundaries. 
We filtered background flares beyond 3$\sigma$ using the light curve filtering script {\tt lc\_clean}. Readout artifacts were subtracted. Point sources were detected in a 0.3--7.0 keV image with {\tt wavdetect}, supplied with a 1.0 keV exposure map. The detection threshold was set to 10$^{-6}$ and the scales of {\tt wavdetect} ranged from 1 to 8, in steps, increasing by a factor of $\sqrt{2}$.

 \begin{deluxetable}{ccccc}

\tablewidth{0pc}
 \centering
\tablecaption{Chandra observational log for the analyses of NGC~1404}
\tablehead{
\colhead{Obs ID}&\colhead{Instrument}&\colhead{exposure (ksec)}&\colhead{RA (deg)}&\colhead{Dec (deg)}}
\startdata
2942&ACIS-S&29.0&54.72&-35.59\\
4174&ACIS-I&45.3&54.71&-35.58\\
4175&ACIS-I&46.0&54.88&-35.76\\
9798&ACIS-S&18.3&54.71&-35.58\\
9799&ACIS-S&21.3&54.71&-35.58\\
16231&ACIS-S&59.9&54.76&-35.59\\
16232&ACIS-S&61.1&54.77&-35.59\\
16233&ACIS-S&98.4&54.77&-35.59\\
16234&ACIS-S&90.3&54.84&-35.65\\
17540&ACIS-S&28.5&54.76&-35.59\\
17541&ACIS-S&24.7&54.76&-35.59\\
17548&ACIS-S&48.0&54.77&-35.59\\
17549&ACIS-S&61.3&54.71&-35.64
\enddata
\tablecomments{The Obs ID of the rest of the observations used in making the mosaic image of the Fornax Cluster are 319, 624, 3949, 4168, 4169, 4170, 4171, 4172, 4173, 4176, 4177, 9530, 13185, 13257, 14527, 14529, 16639.}
\end{deluxetable}

\subsection{\bf Imaging analyses}

Images in 7 energy bands: 0.5--0.7 keV, 0.7--0.9 keV, 0.9--1.1 keV, 1.1--1.3 keV, 1.3--1.5 keV, 1.5--1.7 keV, and 1.7--2.0 keV were generated. We normalized these images with monochromatic exposure maps defined at the central energy of each band. For each image we subtracted an approximation of the background using the blank-sky fields available in the CALDB. The background level was normalized by the count rate in the 9.5 -- 12.0 keV relative to the observation. We replaced point sources with pixel values interpolated from surrounding background regions using {\tt dmfilth}. A final 0.5--2.0 keV image of the Fornax Cluster was produced by adding all these 7 narrow band images as shown in Figure~\ref{fig:fox}.
To maximize the ISM emission over the ambient cluster emission,
we restrict the image analysis to the energy band of 0.7--1.3 keV for the observation of NGC~1404 as shown in Figure~\ref{fig:N1404}.

\subsection{\bf Spectral analysis}

We extracted spectra for the regions of interest on the ACIS-S3 and ACIS-I chips. 
Spectral response matrices were produced for each region with the CIAO tools {\tt mkwarf} and {\tt mkacisrmf}.  All spectra were grouped to have at least one count per energy bin. Spectral fitting was performed with {\sl XSPEC} 12.7 using the C-statistic. The energy range for spectral fitting was restricted to 0.5--7.0 keV. 
We adopted the solar abundance standard of Asplund at al.\ (2006) in thermal spectral models. 
Photoionization cross-sections were from Balucinska-Church \& McCammon (1992). 
We adopted a Galactic 
hydrogen column of $N_{\rm H}=1.5\times10^{20}$ cm$^{-2}$ toward NGC~1404, which was deduced 
from the LAB map (Kalberla et al.\ 2005) incorporated in the {\sl HEASARC}
$N_{\rm H}$ tool. {\tt phabs} was used to model the foreground absorption in all the spectral fitting.
We employed two different kinds of background for specific purposes. We used the blank-sky background when we intend to study the gas properties of the Fornax Cluster gas. 
To measure the gas properties of NGC~1404's gaseous ISM (remnant core plus stripped tail), we extract spectra from regions adjacent to NGC~1404 and applied it as the local background.

\begin{figure*}[h]
   \centering
    \includegraphics[width=0.85\textwidth]{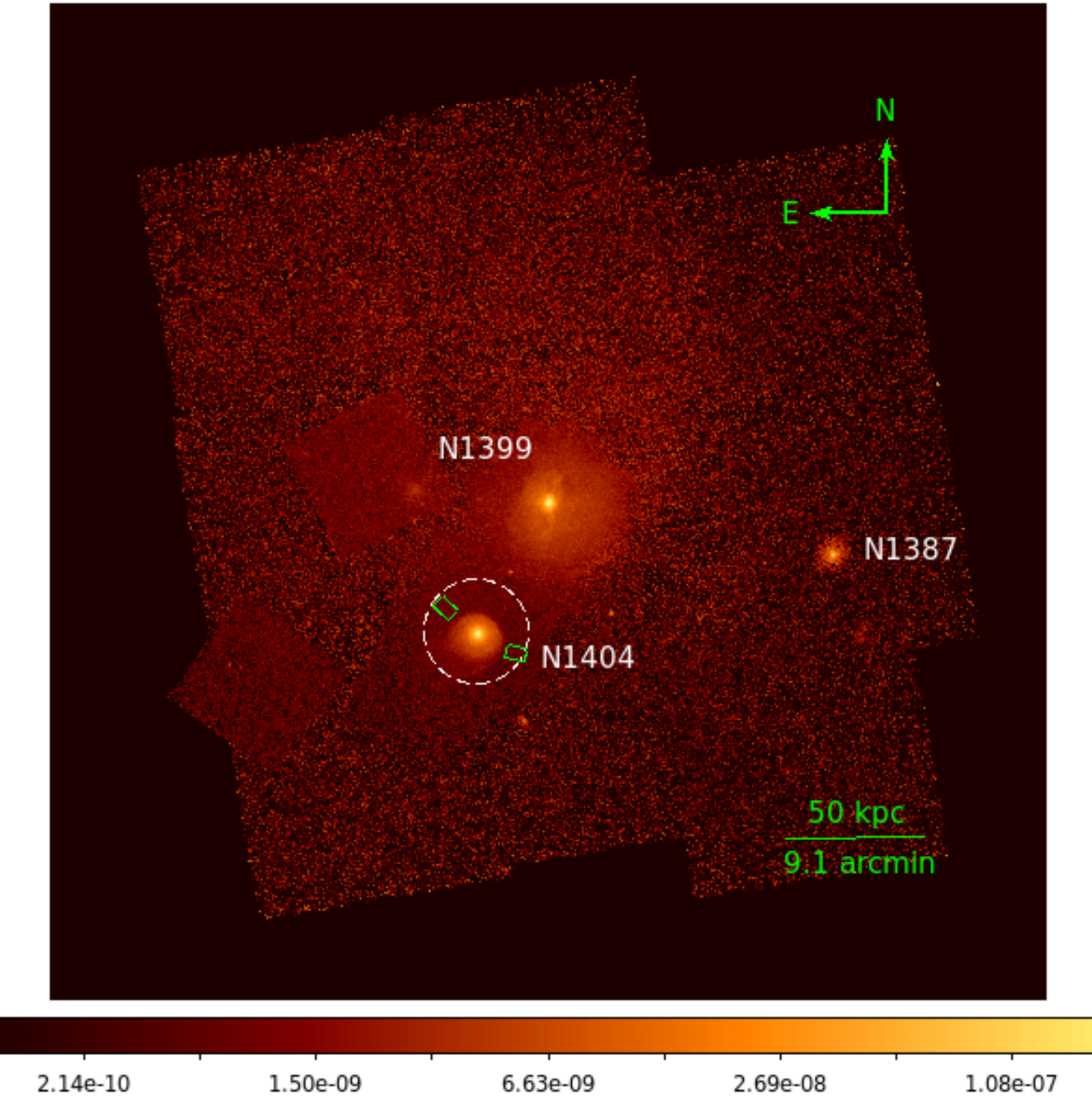}
\figcaption{\label{fig:fox} Mosaic {\sl Chandra} image of the Fornax Cluster in the energy band of 0.5--2.0 keV and in the unit of photon\,cm$^{-2}$\,s$^{-1}$. The image was exposure-corrected with blank-sky background subtracted. White circle indicates the frame of Figure~\ref{fig:N1404}. Green boxes are ``Region ICM" (see text).}
\end{figure*}

\begin{figure*}[h]
   \centering
 %  \hspace{-18.75 mm}
   \includegraphics[width=0.8\textwidth]{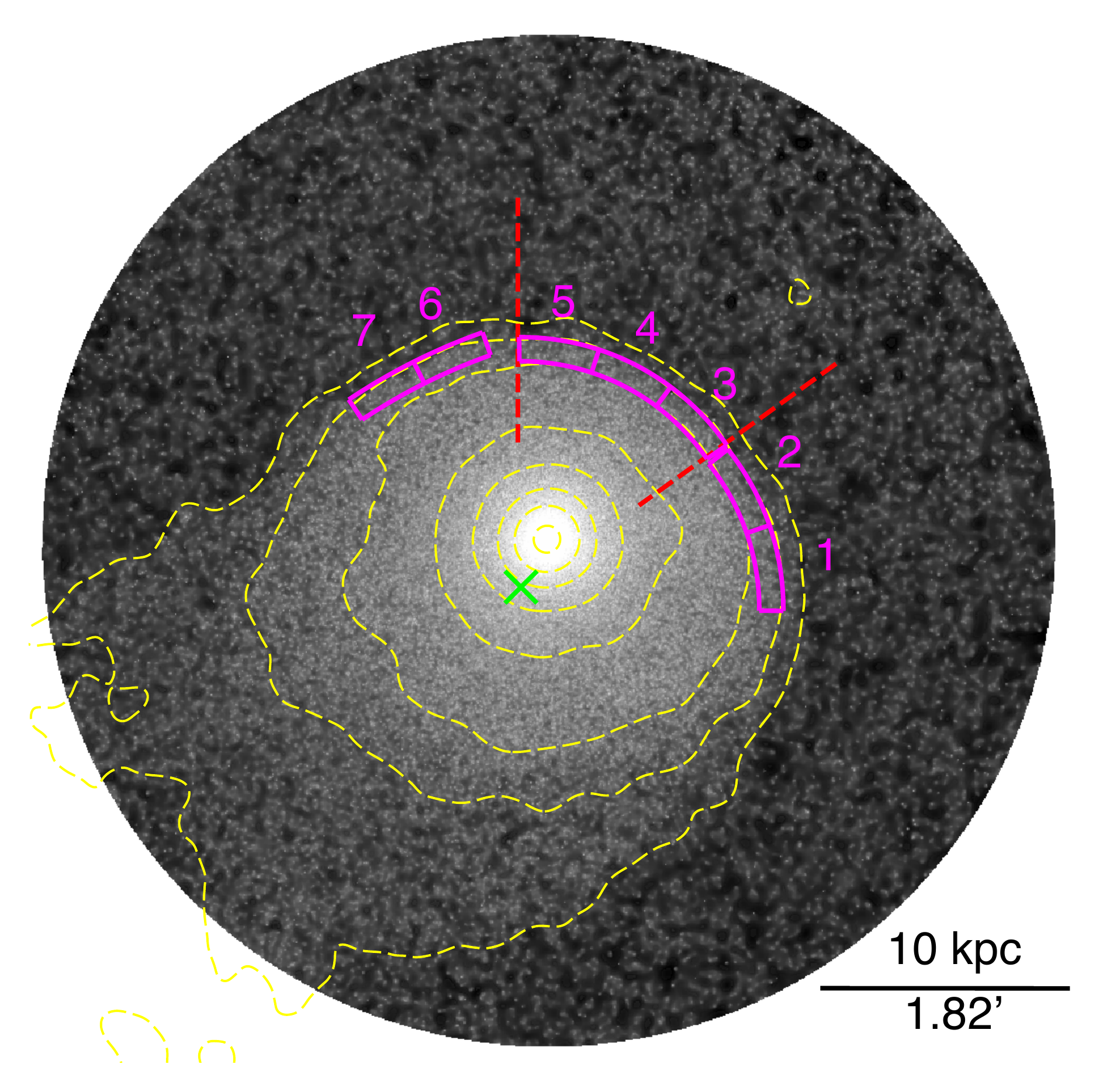}
\figcaption{\label{fig:N1404} {\sl Chandra} image of NGC~1404 in the 0.7--1.3 keV energy band. The image was exposure-corrected with blank-sky and readout background subtracted and point sources removed. We identify the red wedge as the sharpest edge, for which we derived the surface brightness profile (green cross marks the curvature center). 
Regions 1--7 were used to derive the azimuthal pressure variation along the leading edge (see Figure~\ref{fig:view}-top).}
\end{figure*}

We extracted spectra from an annular sector (35$^{\circ}$--90$^{\circ}$) just inside the cold front with a radial width of $15^{\prime\prime}$. This samples the gas properties of the ISM just inside the contact edge. Hereafter, we refer to this region as Region ISM. We fit the spectra with the model ${\tt phabs}\times{\tt vapec}$; local background was applied. The abundances of O, Mg, Si, S, and Fe were allowed to vary freely; all other elements were tied to Fe. We obtain a best-fit of [O,Mg,S,Si,Fe]=[$0.94^{+1.33}_{-0.48}$,$0.23^{+0.39}_{-0.15}$,$0.41^{+0.54}_{-0.26}$, $1.04^{+1.82}_{-1.04}$,$0.52^{+0.52}_{-0.18}$] (also see Table~2). 
To determine the general properties of the ambient Fornax ICM, we extracted spectra from regions near NGC~1404 and at a similar distance to NGC~1399, referred as Region ICM (marked in Figure~\ref{fig:fox}). We fit these spectra to the same model but using blank-sky background. We have to tie O and Mg abundances to constrain the fit. The best-fit result is listed in Table~2. 
When we let $N_{\rm H}$ free to vary in the fit, the best-fit temperature and norm of Region ISM varies by 0.5\% and 1.7\% respectively and those of Region ICM varies by 1.5\% and 6\% respectively.

 \begin{deluxetable}{ccccccc}
\tablewidth{0pc}
 \centering
\tablecaption{Results of spectral analyses}
\tablehead{
\colhead{Region}&\colhead{T (keV)}&\colhead{Fe ($Z_{\odot}$)}&\colhead{$n_e$ (cm$^{-3}$)}&\colhead{${\Lambda}$ (ergs\,cm$^{3}$/s)}}
\startdata
ISM &$0.6{\pm0.02}$&$0.52^{+0.52}_{-0.18}$&0.0061&$8.63\times10^{-24}$\\
 ICM  &$1.57{\pm0.04}$&$0.30^{+0.05}_{-0.04}$&0.0012&$2.60\times10^{-24}$
\enddata

\tablecomments{Both regions were fit to a single {\tt vapec} thermal model. ${\Lambda}$ denotes the X-ray emissivity (cooling function) derived from {\sl XSPEC} for the 0.7--1.3 keV energy band. Errors are at 1$\sigma$ confidence level. The ICM density is obtained from the spectral fit and a beta-model assumption while the ISM density is based on the density jump derived in the surface brightness analysis.}

\end{deluxetable}

%\bigskip

%\bigskip

\section{\bf Results}

\subsection{\bf The ambient ICM properties}

We determine the gas properties of the free stream. 
We fit the surface brightness profile in the 0.5--2.0\,keV energy band of the Fornax Cluster to a single $\beta$ model (leaving out the central region associated with NGC~1399) to obtain $r_c = 38.5\pm3.7^{\prime\prime}$ and $\beta=0.53\pm0.01$.
An equivalent electron density profile can be expressed as 
\begin{equation}
n_{e}(r)=n_{\rm 0}\left[1+(\frac{r}{r_{c}})^2\right]^{-3\beta/2}.
\end{equation}
We calculate $n_{0}$ from the best-fit {\tt norm} of Region ICM as 
$$
{\rm norm} = \frac{10^{-14}}{4\pi[D_A(1+z)]^2}
$$
\begin{equation}
\int_{R^2}^{\infty}{\frac{{n_{\rm 0}}^2}{1.2}\left[1+(\frac{r}{r_{c}})^2\right]^{-3\beta}\frac{S}{\sqrt{r^2-R^2}}{\rm d}r^2}, 
\end{equation}
where $S$ is the area of Region ICM and $R$ is its projected distance from NGC~1399.
We determine the ICM electron density near NGC~1404 to be $1.20\pm0.15 \times 10^{-3}$\,cm$^{-3}$ (Table~2). We use this density and a temperature of k$T = 1.57\pm0.05$ keV for the gas properties of the free stream surrounding NGC~1404. This corresponds to a sound speed of $c_s=\sqrt{\gamma T/m_p\mu}=631$ km\,s$^{-1}$, where $\gamma=5/3$ and the average molecular weight $\mu=0.62$. We refer the ambient ICM as the free stream; its pressure is denoted as $P_1$ below. 

\subsection{\bf Surface brightness and contact discontinuity}

\begin{figure}[h]
   \centering
       \includegraphics[width=0.5\textwidth]{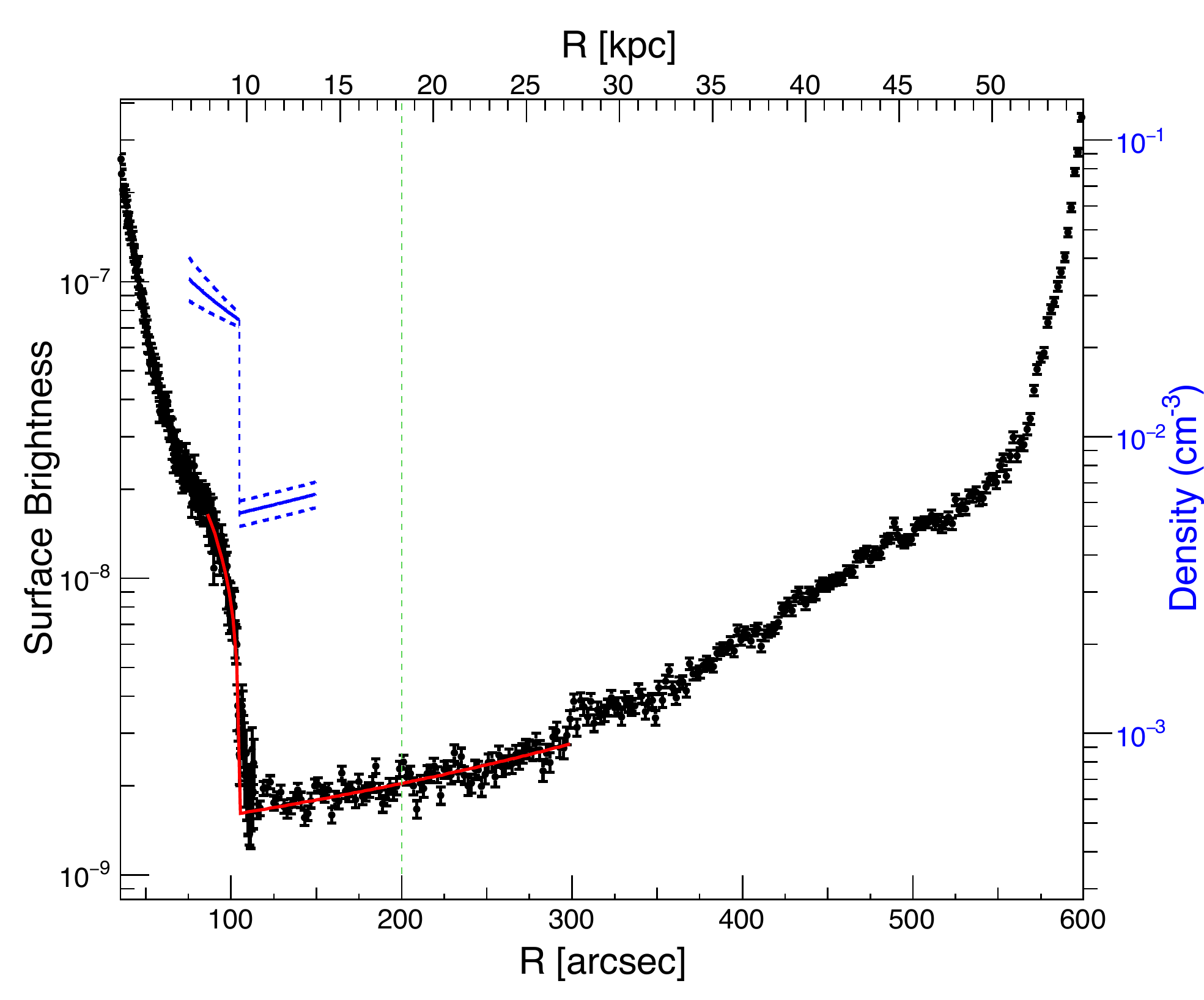}
\figcaption{\label{fig:foxnax} Black dots: surface brightness profile (in units of photons\,s$^{-1}$\,cm$^{-2}$\,arcsec$^{-2}$) from NGC~1404 to NGC~1399 across the leading edge of NGC~1404. Red line: the best-fit of the surface brightness profile consisting of a gaseous ISM component and an ambient ICM component. Blue line: density jump across the leading edge derived from the best-fit. Green dashed line marks the position of the expected shock front.}
\end{figure}

As shown in Figure~\ref{fig:fox}, the atmosphere of NGC~1404 displays a sharp edge facing the cluster center (northwest side of NGC~1404), with a gaseous tail to the southeast, consistent with the stripping scenario.
We identify the leading edge of NGC~1404 as its sharpest edge, for which we obtained a surface brightness profile in the sector from 35$^{\circ}$--90$^{\circ}$\footnote{All angles are measured counter clockwise from west.} centered on the curvature center of the front (see Figure~\ref{fig:N1404}). 
We also included surface brightness profile derived from near NGC~1399 to the leading edge of NGC~1404 as shown in Figure~\ref{fig:foxnax}.
We expect the profile just outside the edge to follow the surface brightness of the Fornax Cluster in the form of a $\beta$ model. 
%$$
%S_{\rm ICM}(r)=S_0 \left[1+(\frac{610-r}{r_c})^2\right]^{-3\beta+0.5}, 
%$$
%here $r$ is relative to the center of the curvature; $610^{\prime\prime}$ is the distance between the center of the curvature and NGC~1399. This then gives 
%The emissivity of the external gas can be obtained through: 
%$$\epsilon_{\rm ICM}(r) =-\frac{1}{2\pi r}\frac{d}{dr}\int_{r^2}^{\infty}\frac{S_{\rm ICM}(R)dR^2}{\sqrt{R^2-r^2}},$$
%where $r$ is the distance relative to NGC~1399.
We assume the gaseous ISM just inside the edge can be described by a spherically symmetric plasma of constant temperature and abundance and follows a power-law density distribution:
\begin{equation}
n_{0}(r<r_{\rm edge})=n_{\rm ISM}\left(\frac{r}{r_{\rm edge}}\right)^{-\alpha},
\end{equation}
where $r$ is relative to the center of the curvature.
% such that $r_0=610^{\prime\prime}-r$ and $r_{\rm edge}$ is the position of the leading edge. 
Its corresponding surface brightness profile $S_{\rm ISM}$ takes the form of equation (A4) in Vikhlinin et al.\ (2001).
%$$
%S_{\rm ISM}(r_0<r_{\rm edge})=2r_{\rm edge}\epsilon_{\rm ISM}
%$$
%\begin{equation}
%\left[1-(\frac{r_0}{r_{\rm edge}})^2\right]\left(\frac{r_0}{r_{\rm edge}})^{-3.45\beta_0}\right),
%\end{equation}
%where $\epsilon_{\rm ISM}$ is the emissivity just inside the edge. 

We fit the surface brightness profile of Figure~\ref{fig:foxnax} to a model containing both the Fornax ICM component ($S_{\rm ICM}$) and the NGC~1404 ISM component ($S_{\rm ISM}$ $[r< r_{\rm edge}]$). 
The abrupt density jump $J$ at the boundary, $r_{\rm edge}$, can be related to the observed surface brightness discontinuity 
%by
%$$
%J^2=\frac{S_{\rm ISM}+S_{\rm ICM}}{S_{\rm ICM}}=\frac{2r_{\rm edge}\Lambda_{\rm ISM} {n_{\rm ISM}}^2}{\left(\int_{b^2}^{\infty}\frac{\Lambda_{\rm ICM} {n_{\rm ICM}}^2}{\sqrt{r^2-b^2}}dr^2\right)}+1,
%$$
%where $b=610^{\prime\prime}-r_{\rm edge}$.
$$
J^2=\frac{S_{\rm ISM} +S_{\rm ICM}}{S_{\rm ICM}}=f(\Lambda_{\rm ICM} {n_{\rm ICM}}^2, \Lambda_{\rm ISM} {n_{\rm ISM}}^2).
$$
%where $b=610^{\prime\prime}-r_{\rm edge}$.
%where $n_{\rm ICM}$ is the ICM density at $r=610^{\prime\prime}-r_{\rm edge}$ in equation (1).
The values of the X-ray emissivity $\Lambda$ are listed in Table~2.
%the ratio of $\epsilon_{\rm ISM}$ and $\epsilon_{\rm ICM}$:$$\frac{n_{\rm ISM}}{n_{\rm ICM}}=\sqrt{\frac{\epsilon_{\rm ISM}\Lambda_{\rm ICM}}{\epsilon_{\rm ICM}\Lambda_{\rm ISM}}}$$.
The best-fit model is indicated in the red solid line in Figure~\ref{fig:foxnax}.
A density jump of $5.2\pm0.2$ at $r_{\rm edge}=104\farcs25\pm0.02$ is given by the best-fit model. Assuming a uniform interior density profile, we obtain a density jump of $5.5\pm0.1$ at $r_{\rm edge}=103\farcs85\pm0.06$ instead.

%\newpage

 \begin{deluxetable}{clllll}
\tablewidth{0pc}
 \centering
\tablecaption{Azimuthal variation of gas properties at the leading edge}
\tablehead{\colhead{Region}&\colhead{T (keV)}&\colhead{$n_e$ (cm$^{-3}$)}&\colhead{$\theta$}&\colhead{$P^{\prime}/P_1$}&\colhead{$\theta^{\prime}$}}
\startdata
1 &$0.60^{+0.04}_{-0.04}$&5.8$^{+0.2}_{-0.2}\times10^{-3}$&-38&$1.86\pm0.20$&50\\
2  &$0.66^{+0.04}_{-0.04}$&6.0$^{+0.2}_{-0.2}\times10^{-3}$&-19&$2.12\pm0.20$&39\\
3  &$0.64^{+0.04}_{-0.03}$&6.6$^{+0.2}_{-0.2}\times10^{-3}$&0&$2.25\pm0.19$&35\\
4  &$0.61^{+0.04}_{-0.03}$&6.5$^{+0.2}_{-0.2}\times10^{-3}$&18&$2.11\pm0.19$&39\\
5  &$0.61^{+0.04}_{-0.04}$&6.5$^{+0.2}_{-0.2}\times10^{-3}$&37&$2.12\pm0.20$&49\\
6  &$0.54^{+0.03}_{-0.06}$&5.3$^{+0.3}_{-0.2}\times10^{-3}$&58&$1.52\pm0.22$&64\\
7  &$0.60^{+0.04}_{-0.04}$&5.0$^{+0.2}_{-0.2}\times10^{-3}$&76&$1.60\pm0.20$&79
\enddata
\tablecomments{$\theta$ and $\theta^{\prime}$ are the angles of each region relative to the apparent and actual stagnation point respectively.}
\end{deluxetable}
%\bigskip

\section {\bf Discussion}

ICM, at intermediate radii, can be approximated as an ideal non-magnetized fluid that has smoothly distributed thermal pressure.    
Here, we apply our knowledge of fluid dynamics to the observational results to determine the three dimensional motion and merging history of NGC~1404.

\subsection{\bf The infall geometry}

\begin{figure}[h]
   \centering
       \includegraphics[width=0.45\textwidth]{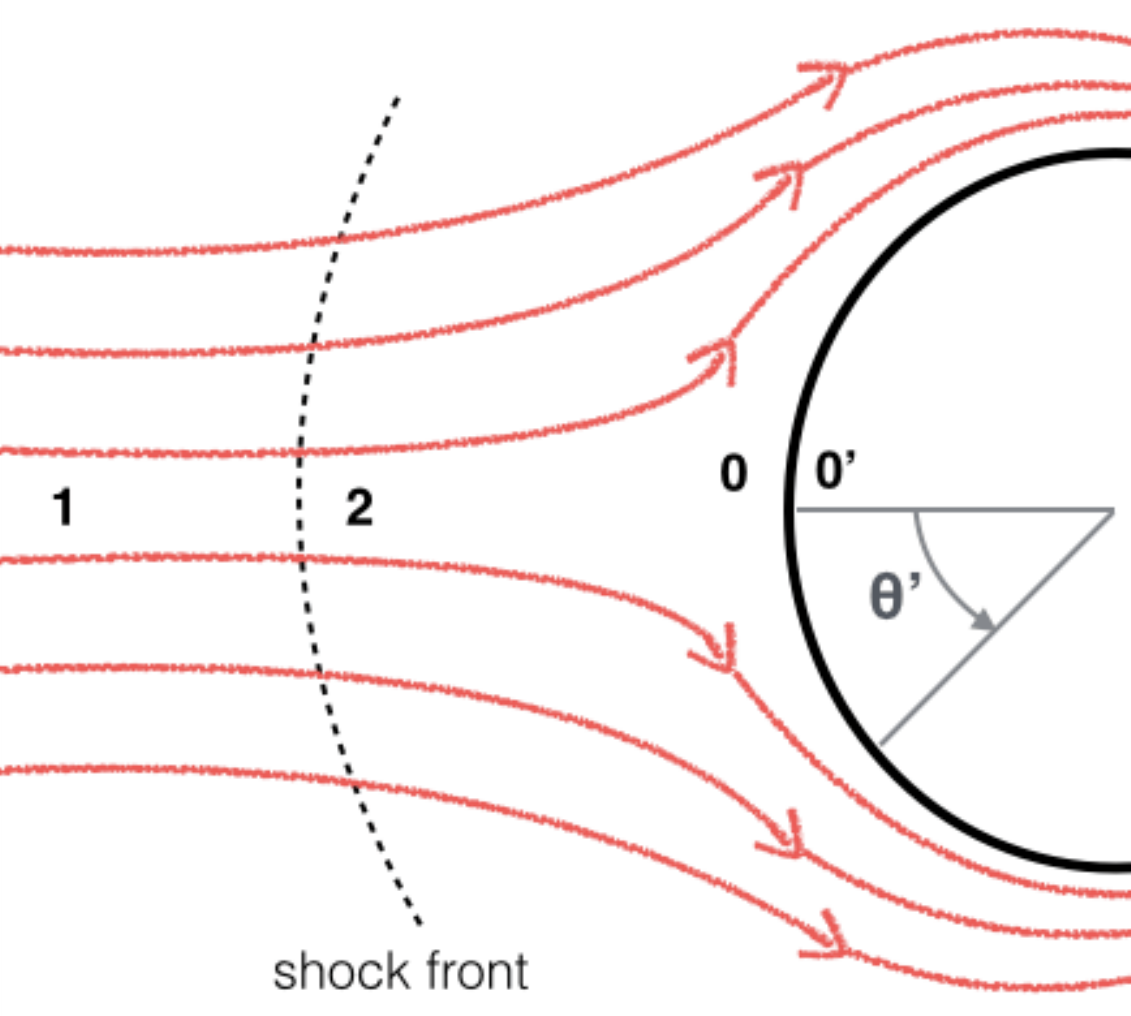}
%\vspace{-0.3cm}
\figcaption{\label{fig:flow} Scheme of flow past a spheroid. Zones 0, 1, and 2 are the stagnation point, the undisturbed free stream, and a (possible) post shock region, respectively. Zone 0' is within the body. $\theta^{\prime}$ is the angle to the actual stagnation point.}
\end{figure}

We consider a simple case of stationary flow of a fluid around a solid sphere as shown in Figure~\ref{fig:flow}, following the scheme in Vikhlinin et al.\ (2001). The fluid pressure distribution along the surface of the sphere varies characteristically with distance from the stagnation point (SP). As a consequence of the Bernoulli principle, the pressure variation along the front can be related to the kinematics of the flow. At the SP, the most upstream point, the static pressure is highest as the local flow velocity is zero. The well-known ram pressure is characterized by the difference between this SP pressure and the free-stream pressure. The fluid pressure at the sphere's surface then decreases with increasing distance from the SP all the way to the sides of the sphere. 

The infalling of NGC 1404 is a very similar scenario, where NGC 1404's atmosphere replaces the solid sphere, and the Fornax ICM flow around (the moving) NGC 1404 constitutes the ambient flow. Thus, if we could measure the Fornax ICM pressure distribution immediately outside the atmosphere of NGC 1404, we could deduce the flow geometry and, thus the 3D motion of NGC 1404.
Directly measuring the ICM pressure just outside the contact discontinuity is difficult because of the low surface brightness of the ICM. However, pressure should be continuous across the contact discontinuity of a cold front (Markevitch \& Vikhlinin 2007). 
The pressure of the NGC 1404 atmosphere just inside the contact discontinuity should be the same as that just outside and the former is much easier to measure thanks to the higher X-ray brightness of the ISM.

We seek to determine the gas properties and their azimuthal variation just inside the cold front.
We extract spectra from 7 regions along the cold front (as shown in Figure~\ref{fig:N1404}-top) chosen to match the curvature of the leading edge. 
Each region has a radial width of 10$^{\prime\prime}$ and an open angle of $\sim18^{\circ}$.
We fit these spectra to a single {\tt vapec} model. We are unable to constrain metallicities, therefore we fix the value of each element to the best-fit of Region ISM. The temperature and normalization of each region are listed in Table~3.
The electron density is derived from the best-fit {\tt normalization} of the {\tt vapec} model\footnote{volume of each small region is calculated as $V=\frac{4}{3}\pi ({r_{\rm out}}^2-{r_{\rm in}}^2)^{3/2}\frac{\Delta \theta}{2\pi}$.}.
The average density of regions 3--5 is 5.2 times that of the ambient ICM, in agreement with the density jump we obtain through the surface brightness modeling.

\begin{figure}[h]
   \centering
       \includegraphics[width=0.5\textwidth]{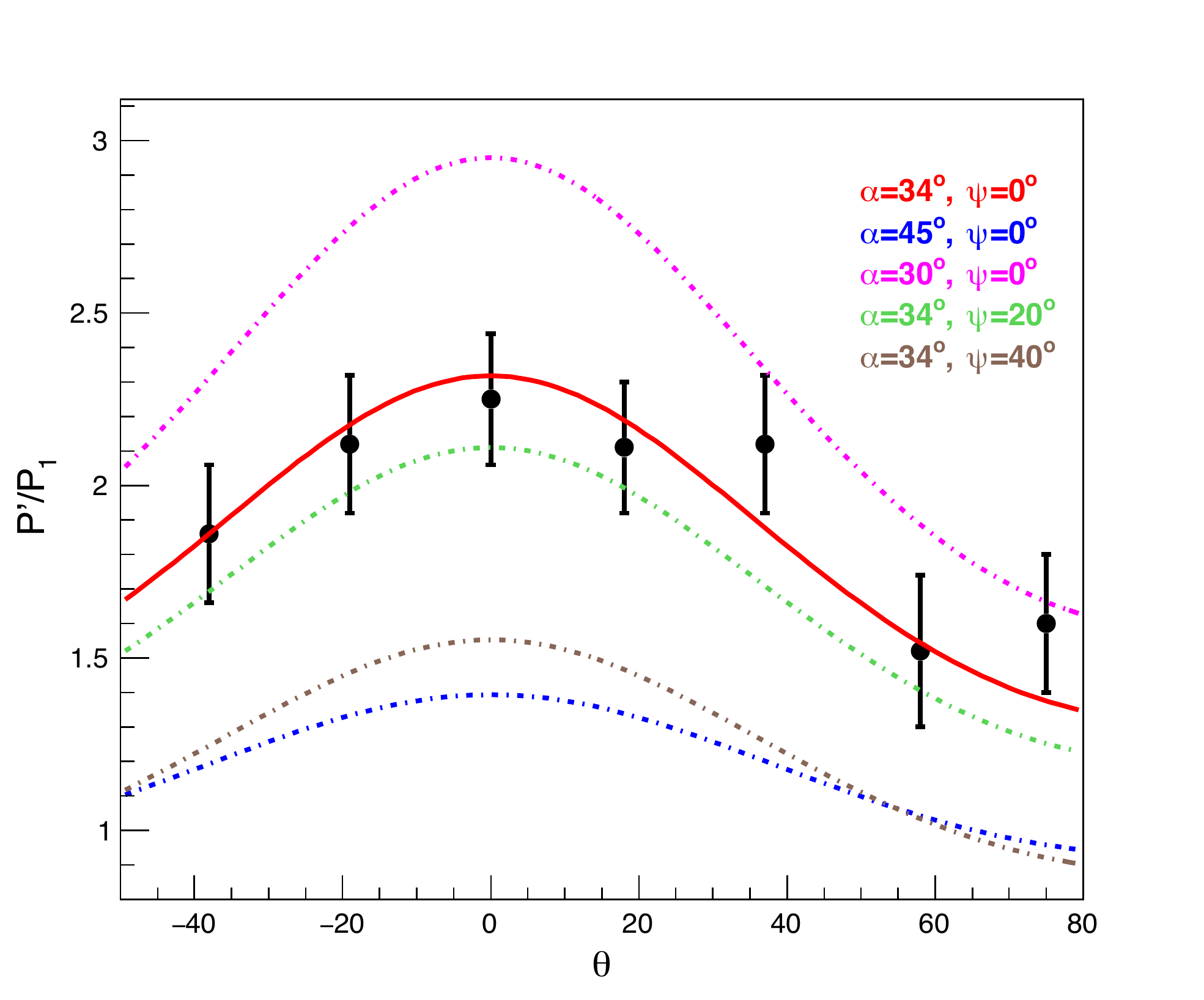}
    \includegraphics[width=0.45\textwidth]{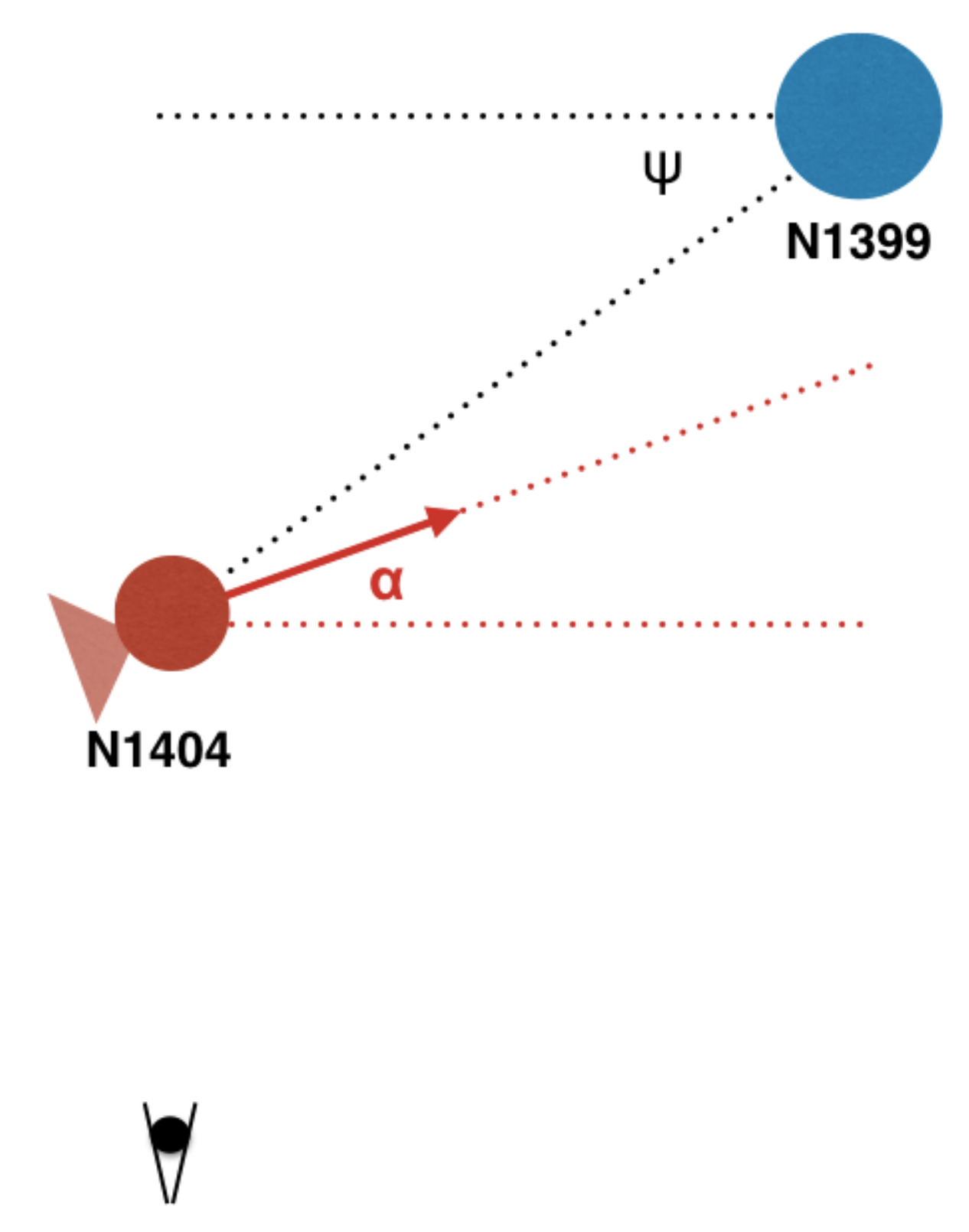}
%\vspace{-0.3cm}
\figcaption{\label{fig:view} {\it top:} angular pressure distribution just inside the front relative to the free stream, corresponding to Regions 1--7 in Figure~\ref{fig:N1404}; $\theta$ equals 0 at Region 3 and increases counter-clock wise. {\it bottom:} illustration of the definitions of $\alpha$ and $\psi$, the inclination angle and the position angle of NGC~1404 respectively.}
% Locations of G1 and G2 are indicated. The peak of cluster gas emission is indicated by cyan cross.}
\end{figure}

%The pressure jump at the stagnation point (SP) is determined by the infalling velocity of NGC~1404. 
The ratio of the pressure at the SP to that in the free stream is determined by the infalling velocity of NGC~1404. 
The Bernoulli equation requires that the pressure declines as the flow velocity increases away from the SP. 
The ratio of the pressure to that in the free stream for each of the 7 regions is listed in Table~3 and plotted in Figure~\ref{fig:view} (top). We assume the pressure, $P^{\prime}$, just inside the front equals that just outside and we refer to the pressure of the free stream as $P_1$.  %Figure~?? represents the variation. 
We take the direction of the largest pressure jump as the direction NGC~1404 is heading in projection. This is region 3, for which we set $\theta=0$. Region 3 also lies opposite to the downstream tail and towards the Fornax center, supporting that Region 3 is in the projected direction of motion. 

For flows with free stream Mach numbers just in
excess of unity,
Vikhlinin et al.\ (2001) use ${\cal M}' =
1.1 \sin \theta'$ to approximate the Mach number at the surface of the sphere, where $\theta'$ is the angle to the SP (Figure~\ref{fig:flow}). This relation is based on the simulations of Rizzi (1980) for a free stream Mach
number of ${\cal M}_1 = 1.05$.  This solution is also applicable for slightly
greater ${\cal M}_1$ according to laboratory measurements (Heberle et al. 1950).
Furthermore, it has the same angular dependence as the flow velocity
for irrotational, inviscid flow past a sphere. 
For this study, we assume that the Mach number at the surface of the sphere takes the form of ${\cal
M}' = A \sin \theta'$, where $A$ is a model parameter.

NGC~1404 has a line-of-sight velocity of $v_{\rm L}=454$ km\,s$^{-1}$ relative to the average of the member galaxies in the Fornax Cluster (Drinkwater et al.\ 2001) ($v_{\rm L}=522$ km\,s$^{-1}$ relative to NGC~1399). This is comparable to the sound speed.
It is highly unlikely that the velocity of NGC~1404 relative to the ICM is perpendicular to our line-of-sight, so that we do not have a direct measurement of the pressure at the SP.
The pressure of the ambient free stream also depends on the distance of NGC~1404 relative to the plane of the sky that contains the cluster center.
 
To take all these factors into account, we construct a model 
that relates the pressure variation along the front to the infall dynamics and geometry of NGC~1404.
%for the pressure variation over the front between the ICM and the ISM in NGC~1404 and its ratio to the pressure in the free stream (see Appendix). 
There are three parameters in the model: $\alpha$, the inclination of the orbital velocity of NGC~1404 to the plane of the sky,
$\psi$, the inclination of the vector from NGC~1404 to the cluster center from the plane of the sky (Figure~\ref{fig:view}-bottom), and $A$, where the Mach number along the front is $\mathcal{M}^{\prime}=A\sin\theta^{\prime}$. The detailed form of the model and its derivation are described in Appendix. 
The infall speed (Mach number) is determined by the value of $\alpha$, since $\mathcal{M}={\cal M}_1=v_L/c_s/{\rm sin \alpha}$, which decreases as both $\alpha$ and $\psi$. 
The best-fit (red solid line in Figure~\ref{fig:view} - top) indicates that NGC~1404 is near the plane of the sky containing the cluster center ($\psi=0.02^{\circ}\pm28.8$) and it is falling through the ICM at ${32.9^{\circ}}^{+0.8}_{-3.6}$ with an infall velocity of 830 km\,s$^{-1}$ ($\mathcal{M}_1=1.32^{+0.15}_{-0.03}$).

\subsection{\bf Search for shock features}

\begin{figure}[h]
   \centering
    \includegraphics[width=0.45\textwidth]{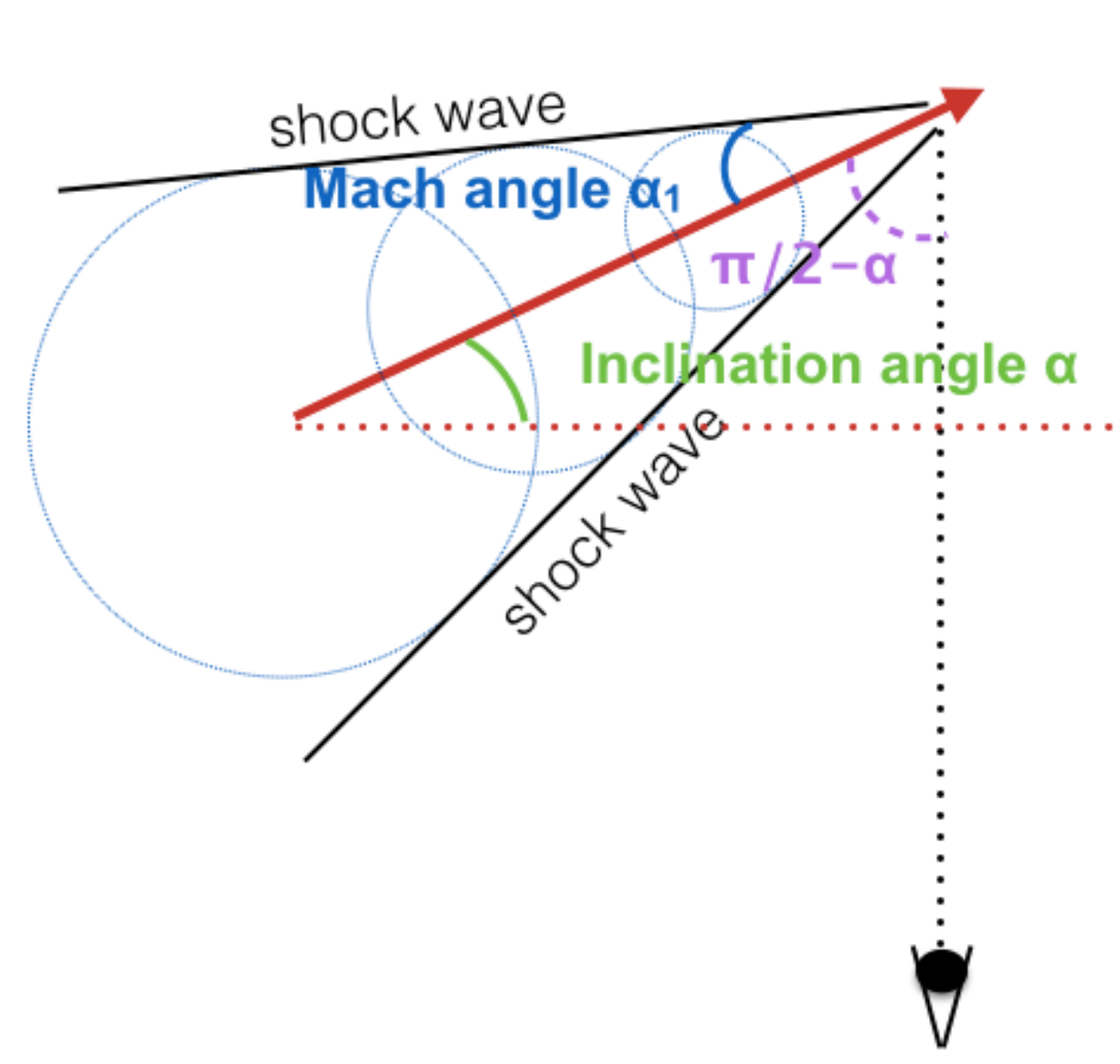}
       \includegraphics[width=0.5\textwidth]{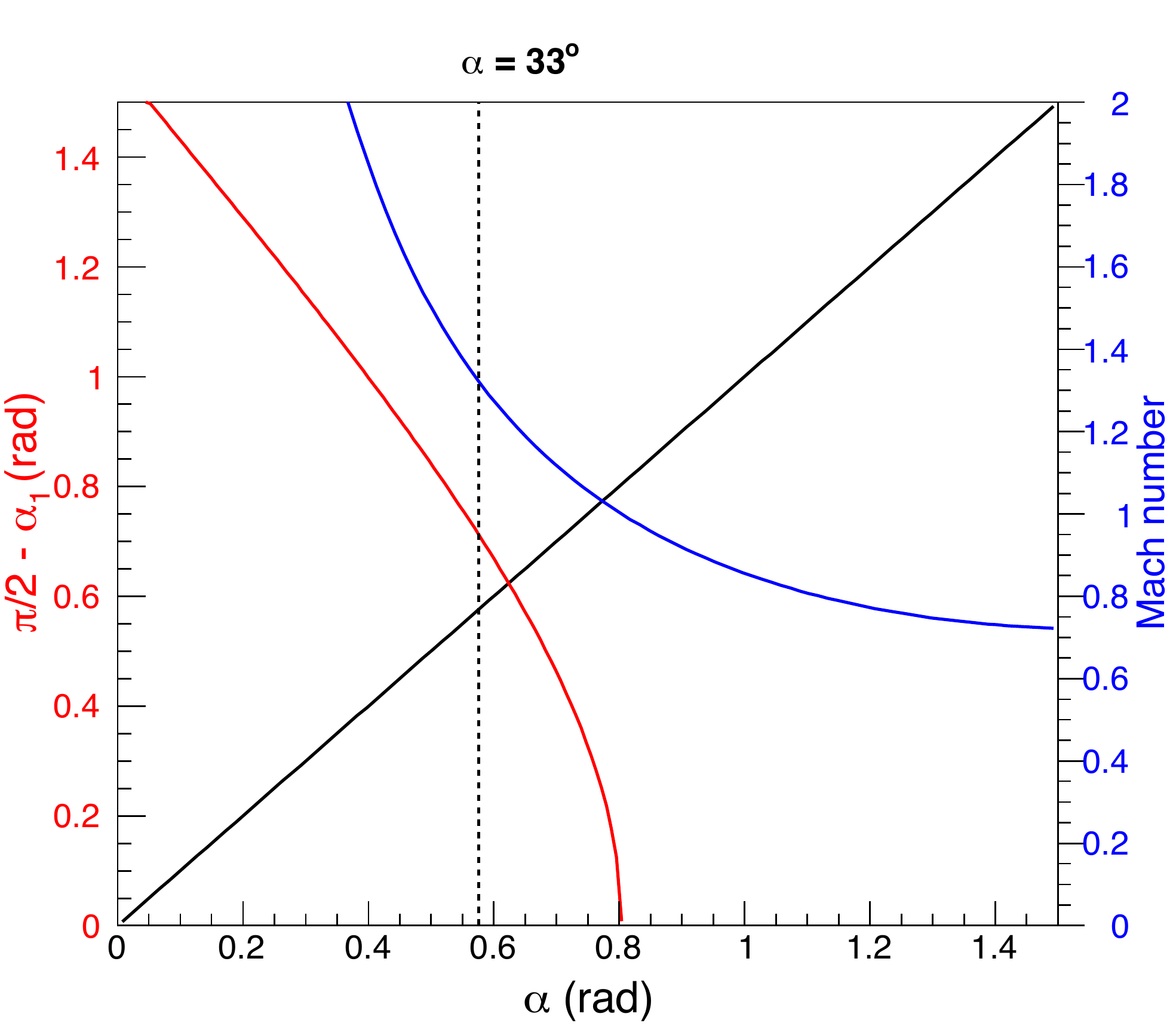}
%\vspace{-0.3cm}
\figcaption{\label{fig:mach} {\it Top:} Scheme of a shock wave and our point of view. For a shock feature to be detected, our line-of-sight needs to be outside the Mach cone, requiring $\frac{\pi}{2}-\alpha > \alpha_1$. {\it Bottom:} $(\frac{\pi}{2}-\alpha_1)$ and Mach number as a function of $\alpha$. $\alpha_1$ is the Mach angle and $\alpha$ is the inclination angle of NGC~1404. Black solid line represents $\alpha=\frac{\pi}{2}-\alpha_1$. Shock features are detectable for cases above the black solid line ($\alpha+\alpha_1 < \frac{\pi}{2}$). Black dashed line indicates our best-fit of $\alpha=33^{\circ}$.}
% Locations of G1 and G2 are indicated. The peak of cluster gas emission is indicated by cyan cross.}
\end{figure}

For an infall velocity of $\mathcal{M}$=1.32, a density jump of 1.5 and a temperature jump of 1.3 (corresponding to 2 keV) are expected at a possible bow shock based on Rankine-Hugoniot shock equations (Laudau \& Lifshitz 1959)
%$$
%\frac{P_0}{P_1}=\left(\frac{\gamma+1}{2}\right)^{(\gamma+1)/(\gamma-1)}\mathcal{M}^2\left[\gamma-\frac{\gamma-1}{2\mathcal{M}^2}\right]^{-1/(\gamma-1)}  $$
\begin{equation}
\frac{\rho_2}{\rho_1}=\frac{(1+\gamma)\mathcal{M}^2}{2+(\gamma-1)\mathcal{M}^2},
\end{equation}
\begin{equation}
\frac{T_2}{T_1}=\left[\frac{2\gamma\mathcal{M}^2-(\gamma-1)}{(\gamma+1)^2}\right]\left[\gamma-1+\frac{2}{\mathcal{M}^2}\right].
%\frac{T_0}{T_1}=1+\frac{\gamma-1}{2}\mathcal{M}^2.
\end{equation}
%Such a shock heated cluster gas of 2 keV may not be visible if the angle of the Mach cone is smaller than the inclination angle. 
According to Farris \& Russell (1994), the distance between the leading edge and the bow shock for an impenetrable obstacle can be approximated as 
\begin{equation}
D_{CS}=0.8R\frac{(\gamma-1)\mathcal{M}^2+2}{(\gamma+1)(\mathcal{M}^2-1),}
\end{equation}
where $R$ is the radius of a nearly spherical body.  
$D_{CS}$ would be 11 kpc for $\mathcal{M}$=1.32. At an inclination angle of $33^{\circ}$, we expect the bow shock to lie 9 kpc in projection ahead of the leading edge (marked in Figure~\ref{fig:foxnax}). 
No obvious enhancement is observed in the surface brightness distribution. 
We do note $\sim$\,2 keV gas close to the expected location in the temperature map but this is consistent with ICM temperature fluctuation (Scharf et al.\ 2005; Murakami et al.\ 2011). Note that Equation~(6) is for collisionless bow shocks in incompressible flows, not entirely applicable to NGC~1404. Numerical simulation for infalling early-type galaxies suggests that its shock feature should lie beyond the $D_{CS}$ estimated here (Roediger et al.\ 2015).   
On the other hand, 
for a shock feature to be detected, our line-of-sight needs to be outside the Mach cone (Figure~\ref{fig:mach}-top). This requires the sum of the Mach angle ($\alpha_1$) and the inclination angle ($\alpha$) to be smaller than $\frac{\pi}{2}$. In Figure~\ref{fig:mach}-bottom, we show $(\frac{\pi}{2}-\alpha_1)$ and Mach number as a function of $\alpha$. In this case, the Mach angle is $\alpha_1$ = sin$^{-1}\frac{1}{\mathcal{M}}=50^{\circ}$ and $(\alpha+\alpha_1)$ is marginally smaller than $\frac{\pi}{2}$. The detection of shock features near NGC~1404 can be challenging.

%We note that directions at $\theta=??$ and ?? are not inconsistency as our best-fit model.

\subsection{\bf Merging history of NGC~1404}

Several lines of evidence suggest that NGC~1404 has interacted with NGC~1399 before. 
NGC~1399 contains more metal-poor globular clusters distributed at its outskirts than other typical early-type galaxies, possibly acquired from NGC~1404. 
For a bright cluster elliptical, the specific frequency of globular clusters in 
NGC~1404 is exceptionally low, indicating that it has lost some of its globular clusters (Forbes et al.\ 1997b; Forbes et al.\ 1998; Grillmair et al.\ 1994; Minniti et al.\ 1998; Kissler-Patig et al.\ 1999).  Bekki et al.\ (2003) reproduced the globular cluster distribution for the NGC~1399/1404 complex through a tidal stripping simulation. 
Similar conclusions can be drawn from observations of planetary nebula. Napolitano et al.\ (2002) found a disturbed velocity structure of NGC~1399 showing a peak rotation that is 12.6 kpc offset from the optical center. They suggest that NGC~1404 has undergone a flyby of NGC~1399.

Independent evidence is provided by the diffuse hot gas content of NGC~1399 which displays obvious asymmetry and sloshing structure (Figure~\ref{fig:fox}), implying that NGC~1399 was disturbed by a sizable object. In the field of the Fornax Cluster, the most likely suspect is NGC~1404.  
Thus, this is unlikely to be NGC~1404's first approach to the inner region of the Fornax ICM. 
Note that the dynamic time of a cluster is much longer than that of a galaxy; NGC~1404 should be able to regain its still and spherical atmosphere at its second infall. 
%This would naturally explain why the stripped tail is so short for such a bright early-type galaxy.  
We can also estimate its expected velocity assuming NGC~1404 is falling into this cluster from infinity (first entry). 
The potential energy distribution of the Fornax Cluster can be approached through the NFW mass profile of Navarro et al.\ (1997):
\begin{equation}
\rho=\frac{\rho_c\delta_c}{(r/r_s)(1+r/r_s)^2},
\end{equation}
where $\rho_c(z)=3H(z)^2/8\pi G$, $\delta_c=\frac{200}{3}\frac{c^3}{{\rm ln}(1+c)-c/(1+c)}$.
We take the average dark matter concentration $c=10$ and scale radius $r_s=75$\,kpc of a typical galaxy group from Gastaldello et al.\ (2007).
We find that NGC~1404 should have reached a relative velocity of 1500 km\,s$^{-1}$ at its current position. This greatly exceeds its best-fit velocity of 830 km\,s$^{-1}$. In contrast, its second infall would have started considerably closer to the cluster center and resulted in a smaller velocity. 
By all means this is an order of magnitude estimate; still it supports the scenario 
that this is not NGC~1404's first infall. Our hydrodynamic simulation tailored to the NGC~1404/Fornax complex has adopted more specific and realistic assumptions including the effect of dynamic friction (Sheardown et al.\ in preparation). This would bring together the merging history of NGC~1404 and the sloshing features at the cluster center.

\section {\bf Conclusions}
We analyzed a deep (670\,ksec) {\sl Chandra} observation of NGC~1404, a gas rich galaxy diving through the intracluster medium of the Fornax Cluster. 
By treating the cluster gas as an ideal compressible flow, we develop a method to determine the 3D position and velocity field of a merging substructure.  
We find a hydrodynamic model explaining well the angular pressure variation measured along the cold front of NGC~1404. 
%Through the angular pressure variations along its leading edge, 
The best-fit model implies that NGC~1404 resides in the same plane of the sky as the cluster center; this galaxy is infalling at an inclination angle of $\alpha={32.9^{\circ}}^{+0.8}_{-3.6}$ with a Mach number of $\mathcal{M}_1=1.32^{+0.15}_{-0.03}$ . 
Projection effects may be responsible for our inability to 
detect significant shock features. 
We also infer that this is unlikely to be NGC~1404's first approach to the central region of the Fornax Cluster. 

%The measurement of the Faraday rotation of intracluster medium demonstrated that thermal pressure greatly exceeds magnetic pressure at the intermediate radii of galaxy clusters (Bonafede et al.\ 2013). 
%The successful application of hydrostatic equations in our study is consistent with the such conclusions.\\

%% References section
\section{\bf Acknowledgments}
We acknowledge helpful discussions with Alexey Vikhlinin. 
This work was supported by {\sl Chandra} Awards GO1-12160X and GO2-13125X issued by the
{\sl Chandra} X-ray Observatory Center which is operated by the Smithsonian Astrophysical Observatory under NASA contract NAS8-03060.

\
\appendix
%\center{Azimuthal Variation of Pressure Discontinuity}
\section{Azimuthal Variation of Pressure Discontinuity}

We use Figure~\ref{fig:flow} to illustrate a flow past a sphere in an unmagnetized gas.  We assume that the flow around the sphere is steady
in a frame moving with it.  We wish to determine the ratio of the
pressure near the surface of the sphere to that in the free stream,
$P' / P_1$, and how it varies over the spherical front.  For a
laminar, steady flow, the pressure on the sphere only depends on the
angle, $\theta'$, between the stagnation point and the point of
interest, measured at the center of the sphere.

First consider subsonic flow (when Zone 2 is absent).
Bernoulli's theorem can be applied between the free stream (Zone 1)
and the surface of the sphere

\begin{equation}
\frac{{v^{\prime}}^2}{2}+\frac{\gamma}{\gamma-1}\frac{P^{\prime}}{\rho^{\prime}}=\frac{{v_{1}}^2}{2}+\frac{\gamma}{\gamma-1}\frac{P_{1}}{\rho_{1}},
\end{equation}
where subscript `1' indicates quantities in the
free stream and primes indicate quantities on the surface of the
sphere.  Extracting a factor of the squared sound speed, $\gamma P /
\rho$, from both sides of this equation, gives after a little algebra

\begin{equation}
{c'^2 \over c_s^2}
=\frac{T^{\prime}}{T_1}=\frac{1+\frac{\gamma-1}{2}{\mathcal{M}_1}^2}{1+\frac{\gamma-1}{2}{\mathcal{M}^{\prime}}^2}
\end{equation}
where $c_s$ and $c'$ are the sound speeds
in the free stream and on the surface of the sphere, respectively.
The Mach numbers are defined by ${\cal M}_1 = v_1 / c_s$ in the free stream and ${\cal M}' = v' / c'$ on the surface
of the sphere.
Assuming that the flow is adiabatic then gives 
\begin{equation}
\frac{P^{\prime}}{P_1}={\left(\frac{T^{\prime}}{T_1}\right)}^{\frac{\gamma}{\gamma-1}}=\left(\frac{1+\frac{\gamma-1}{2}{\mathcal{M}_1}^2}{1+\frac{\gamma-1}{2}{\mathcal{M}^{\prime}}^2}\right)^{\frac{\gamma}{\gamma-1}},  {\qquad} {\rm for}~ \mathcal{M}_1  \leq 1.
\end{equation}

For supersonic motion we must first consider the shock jump
between regions 1 and 2, which are (\S89 Landau \& Lifshitz 1959)
 % Let's start over and take Zone 2 into account. Jump conditions across a shock wave give (\S89 Landau \& Lifshitz):

\begin{equation}
%\begin{cases}
\frac{v_2}{v_1}=\frac{n_1}{n_2}=\frac{(\gamma-1){\mathcal{M}_1}^2+2}{(\gamma+1){\mathcal{M}_1}^2}=g({\mathcal{M}_1}),
\end{equation}
\begin{equation}
\frac{P_2}{P_1}=\frac{2\gamma {\mathcal{M}_1}^2-(\gamma-1)}{\gamma+1}=h({\mathcal{M}_1}),
\end{equation}
\begin{equation}
\frac{T_2}{T_1}=g({\mathcal{M}_1})h({\mathcal{M}_1}).
 % \end{cases}
\end{equation}

Thus, the Mach number in Zone 2 is
related to the Mach number in the free stream by ${\cal
M}_2^2 = {\cal M}_1^2 g ({\cal M}_1) / h ({\cal M}_1)$.  Relating
Zones 2 and 0 by Bernoulli's theorem gives

\begin{equation}
\frac{{v^{\prime}}^2}{2}+\frac{\gamma}{\gamma-1}\frac{P^{\prime}}{\rho^{\prime}}=\frac{{v_{2}}^2}{2}+\frac{\gamma}{\gamma-1}\frac{P_{2}}{\rho_{2}}.
\end{equation}

Solving this equation, we obtain

\begin{equation}
{T' \over T_2} = {1 + {\gamma - 1 \over 2} {\cal M}_2^2 \over 1 +
{\gamma - 1 \over 2} {\cal M}'^2} =\frac{1+\frac{\gamma-1}{2}\frac{g(\mathcal{M}_1)}{h(\mathcal{M}_1)}{\mathcal{M}_1}^2}{1+\frac{\gamma-1}{2}{\mathcal{M}^{\prime}}^2}.
\end{equation}

Assuming that the gas is adiabatic after the shock,

\begin{equation}
\frac{P^{\prime}}{P_2}=\left(\frac{T^{\prime}}{T_2}\right)^{\frac{\gamma}{\gamma-1}}=\left(\frac{1+\frac{\gamma-1}{2}\frac{g(\mathcal{M}_1)}{h(\mathcal{M}_1)}{\mathcal{M}_1}^2}{1+\frac{\gamma-1}{2}{\mathcal{M}^{\prime}}^2}\right)^{\frac{\gamma}{\gamma-1}}.
\end{equation}

Therefore, 
\begin{equation}
\frac{P^{\prime}}{P_1}=h(\mathcal{M}_1)\left(\frac{1+\frac{\gamma-1}{2}\frac{g(\mathcal{M}_1)}{h(\mathcal{M}_1)}{\mathcal{M}_1}^2}{1+\frac{\gamma-1}{2}{\mathcal{M}^{\prime}}^2}\right)^{\frac{\gamma}{\gamma-1}},  {\qquad} {\rm for}~ \mathcal{M}_1  > 1.
\end{equation}

Next, we apply A3 and A10 to the infall of NGC~1404.
%\interfootnotelinepenalty=10000

We assume
that the Mach number along the surface of the sphere takes the form of ${\cal
M}' = A \sin \theta'$, where $A$ is a model parameter.
Projected onto the sky, if $\theta$ is the angle we measure
between the apparent leading edge and the position of interest 
along the contact discontinuity between the
ISM of NGC 1404 and the ICM, then the angle between the
position of interest and the actual stagnation point, $\theta'$, is
given by $\cos \theta' = \cos \theta \cos \alpha$, where $\alpha$ is
the inclination angle defined in Figure 6.  The line-of-sight velocity
of NGC 1404 is $v_L = 454$ km\,s$^{-1}$ and the sound speed in the ICM near
NGC 1404 is $c_s = 631$ km\,s$^{-1}$, so fixing the value of $\alpha$ also
determines the Mach number through ${\cal M}_1 = v_L / (c_s \sin
\alpha)$. If $P_{\rm ref}$ is the ICM pressure on the plane of the sky
containing the cluster center at the projected location of NGC 1404,
it is related to the pressure in the free stream near NGC 1404 by
$P_1 = P_{\rm ref} f (\psi)$, where the angle $\psi$ is defined in
Figure 6 and $f(\psi) = [\{1 + r^2 / (r_c \cos \psi)^2 \} 
   / \{ 1 + (r / r_c)^2\}]^{-3 \beta / 2}$, where $r = 8.5'$ is the projected distance between NGC
1399 and the leading edge of NGC 1404, $r_c=38\farcs{5}$ and $\beta=0.53$ are determined through the surface brightness analysis of the Fornax Cluster. We take $\gamma = 5/3$,
appropriate for a monatomic gas. Our result for the pressure
distribution has three free parameters $A$, $\alpha$, and $\psi$
\begin{equation}
\frac{P^{\prime}}{{P_1}} (\theta) = f(\psi) \left(\frac{1+\frac{\gamma-1}{2}(v_L/\rm sin \alpha/c_s)^2}{1+\frac{\gamma-1}{2}A^2(1-{\rm cos^2\theta cos^2\alpha})}\right)^{\frac{\gamma}{\gamma-1}},  {\qquad} {\rm for}~ \mathcal{M}_1  \leq 1
\end{equation}
and 
\begin{equation}
\frac{P^{\prime}}{{P_1}_{\rm cor}}(\theta)=f(\psi)h(v_L/\sin \alpha/c_s)\left(\frac{1+\frac{\gamma-1}{2}\frac{g(v_L/\rm sin \alpha/c_s)}{h(v_L/\sin \alpha/c_s)}(v_L/\sin \alpha/c_s)^2}{1+\frac{\gamma-1}{2}{A^2(1-\rm cos^2\theta cos^2\alpha)}}\right)^{\frac{\gamma}{\gamma-1}}, {\qquad} {\rm for}~ \mathcal{M}_1  > 1
\end{equation}

Fitting the observed pressure distribution, $P' (\theta) / P_1$, to
this function gives the best fit values $A = 0.95$, $\alpha=37^{\circ}$\,($\mathcal{M}_1=1.26$), and $\psi\sim0$ for the motion and location of NGC~1404.  However, based
on the results discussed above, we assume that we must have $A \ge 1.1$
for ${\cal M}_1 \ge 1.05$, so we fix its value at $A = 1.1$.  The best
fit then gives $\alpha={32.9^{\circ}}^{+0.8}_{-3.6}$\,($\mathcal{M}_1=1.32^{+0.15}_{-0.03}$) and $\psi=0.02^{\circ}\pm28.8$. As shown in Figure~\ref{fig:view}, 
our best-fit model is consistent with almost all the data points. If we take $v_L = 522$\,km\,s$^{-1}$ (relative to NGC~1399 instead of the average of the Foxnax member galaxies) for NGC~1404, we obtain best-fits of $\alpha=37.4^{\circ}$\,($\mathcal{M}_1=1.36$) and $\psi\sim0$.

\begin{figure}[h]
   \centering
       \includegraphics[width=0.495\textwidth]{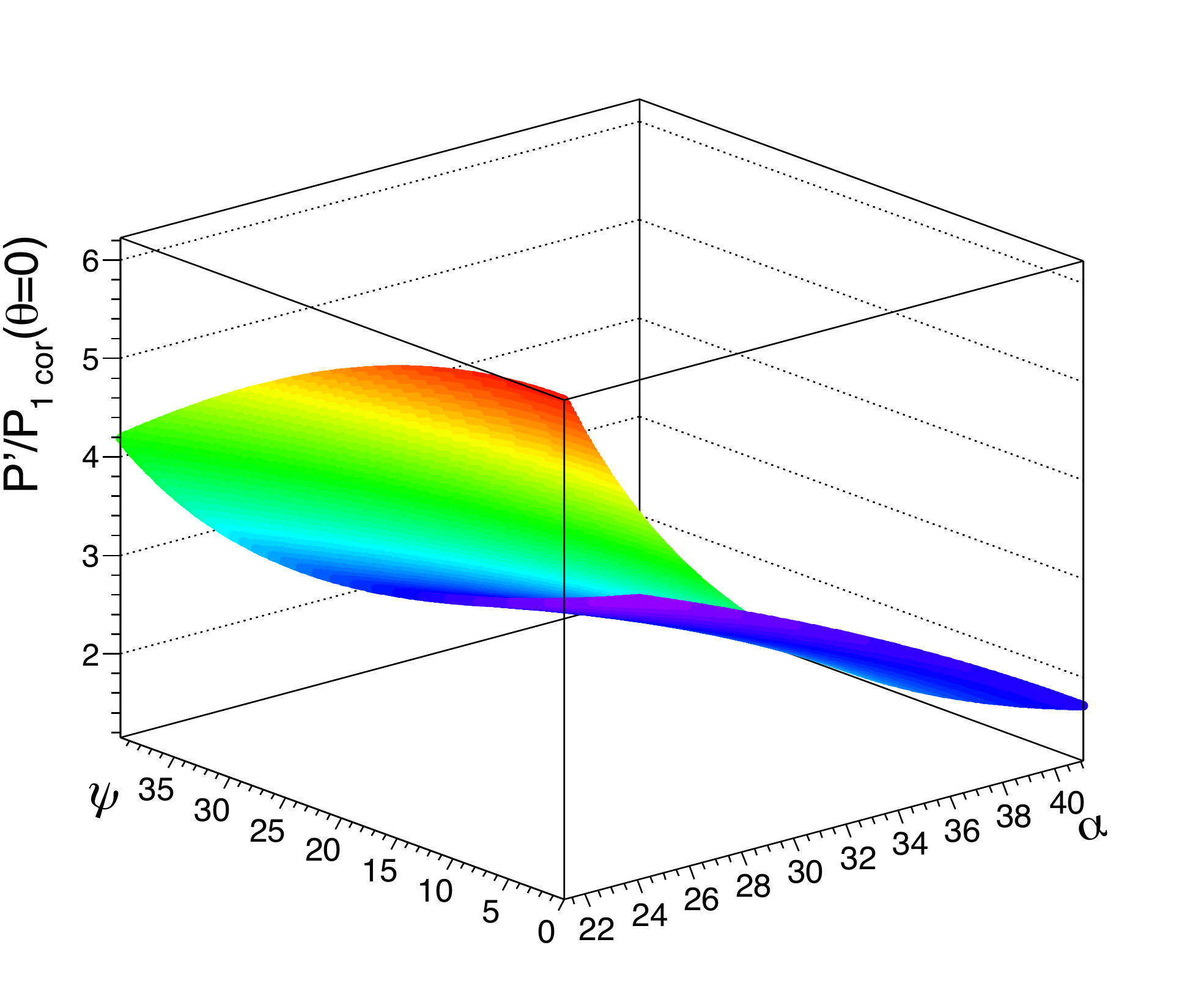}
    \includegraphics[width=0.495\textwidth]{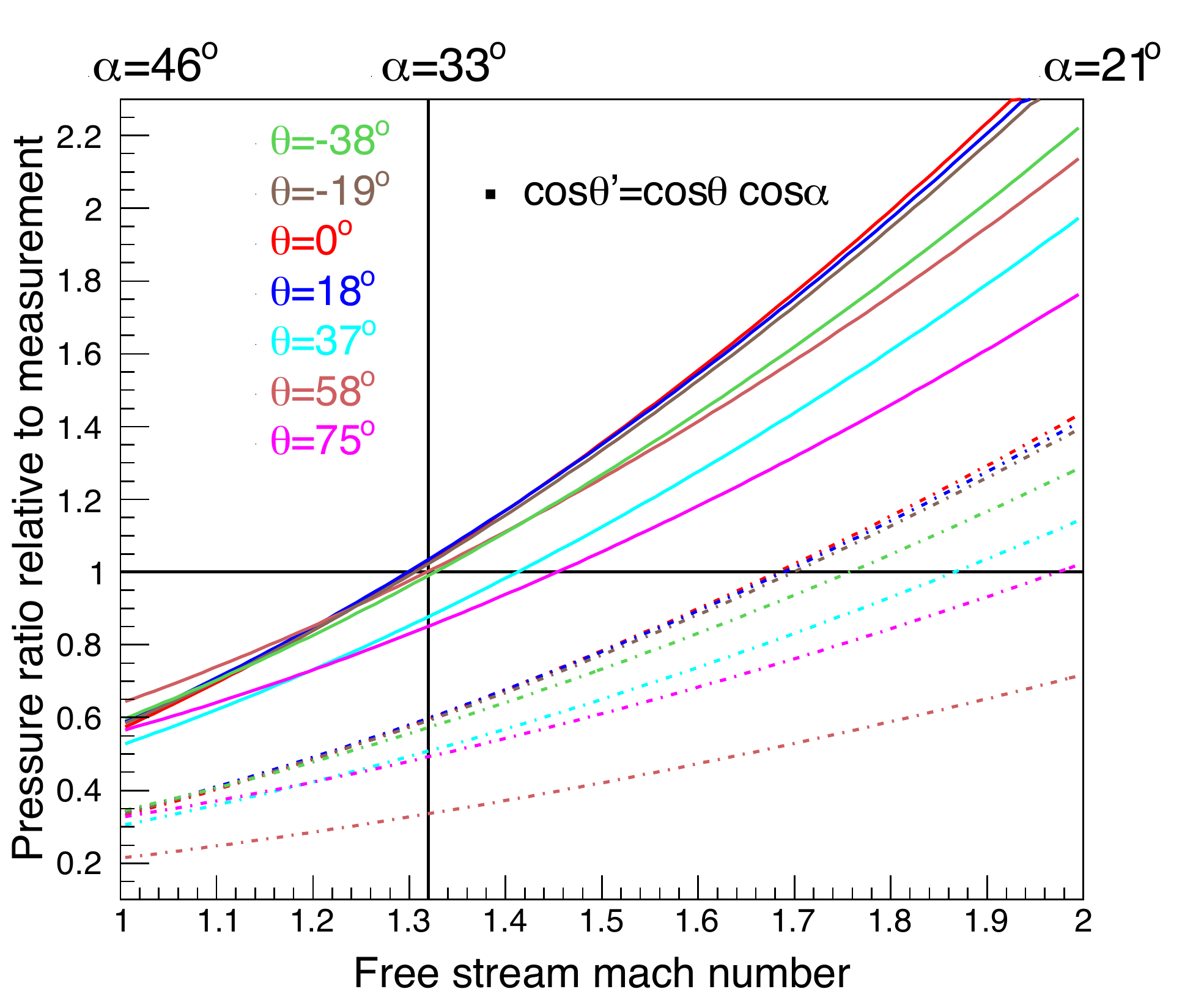}
        \includegraphics[width=0.4\textwidth]{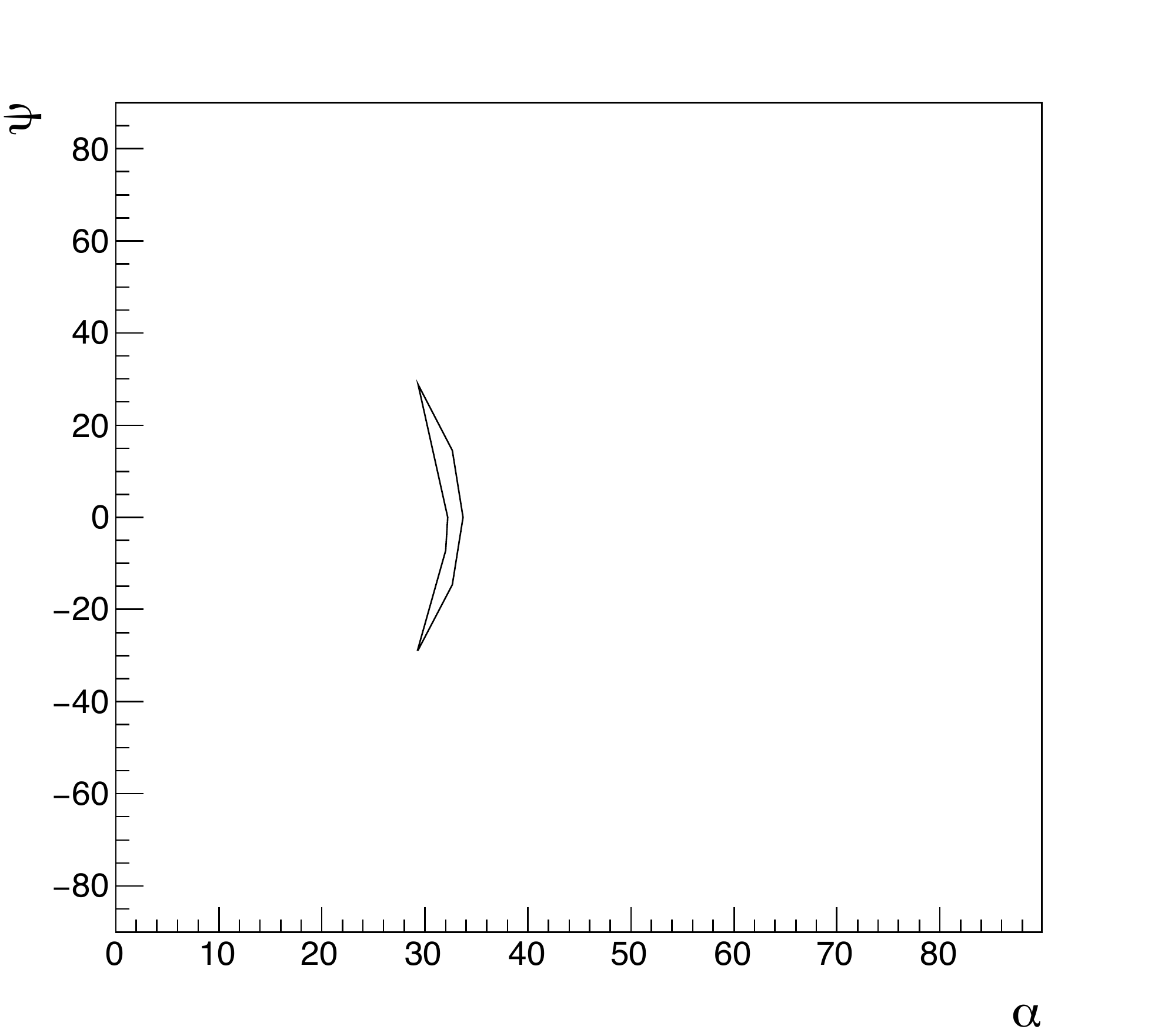}
%\vspace{-0.3cm}
\figcaption{\label{fig:fit} {\it top-left:} expected $\frac{P^{\prime}}{{P_1}}$ as a function of $\psi$ and $\alpha$ when $\theta=0$. {\it top-right:} ratio of expected $\frac{P^{\prime}}{{P_1}}$ and measured $\frac{P^{\prime}}{{P_1}}$ as a function of Mach number (or $\alpha$) for a given $\psi$ (solid lines: $\psi=0$; dashed lines: $\psi=45^{\circ}$). Horizontal black solid line indicates when expectation meets observation. Vertical black solid line marks the best-fit $\alpha=33^{\circ}$. {\it bottom:} $1\sigma$ contour of the parameter space between $\alpha$ and $\psi$.}
\end{figure}

Figure~\ref{fig:fit}(top-left) shows $P' (\theta = 0) / P_1$, at the leading edge of
the ISM as a function of $\psi$ and $\alpha$.  The value is more
sensitive to $\alpha$ than $\psi$.  Figure~\ref{fig:fit} (top-right) shows the ratio
of the model to the observed values of $P'/P_1$ for each data point, as
a function of the Mach number in the free stream, ${\cal M}_1$ (in
effect, $\alpha$), for $\psi = 0$ and $45^\circ$.  Nonzero $\psi$ would
require larger values of ${\cal M}_1$. The parameter space between $\alpha$ and $\psi$ is shown in Figure~\ref{fig:fit} bottom. 

%\newpage

\end{document}